%% file: main.tex
\documentclass[runningheads]{llncs}
\usepackage{graphicx}
\usepackage{comment}
\usepackage[hidelinks]{hyperref}
\usepackage{color}
\usepackage{graphicx,subcaption}
\usepackage{dirtytalk}
\usepackage{graphicx}
\usepackage{amsmath} 
\usepackage{amsfonts} 
\usepackage{amssymb} 
\usepackage{enumitem}
\usepackage{latexsym}
\usepackage{newtxmath}

\begin{document}

\title{TechRank: A Network-Centrality Approach for Informed Cybersecurity-Investment} 

\titlerunning{TechRank}
%

\author{Anita \textsc{Mezzetti}$^{a}$; Dimitri \textsc{Percia David}$^{b}$$^{,}$$^{c}$;  Santiago \textsc{Anton Moreno}$^{d}$; Thomas \textsc{Maillart}$^{b}$;  Michael \textsc{Tsesmelis}$^{c}$
; Alain 
\textsc{Mermoud}$^{c}$}

\authorrunning{A. Mezzetti et al.}
\titlerunning{TechRank}
%
\institute{$^{a}$ EPFL, Section of Financial Engineering\\ $^{b}$ University of Geneva, GSEM, Information Science Institute \\ $^{c}$ Cyber-Defence Campus, armasuisse Science and Technology \\ $^{d}$ EPFL, Mathematics Section}
\maketitle       

\begin{abstract}
\input{chapters/abstract}
\\

\vspace{0.1cm}
\textbf{Acknowledgement}: This research is supported by armasuisse Science and Technology. \\

\noindent
\textbf{Corresponding author}: dimitri.perciadavid@unige.ch
\end{abstract}

\newpage

\section{Introduction}\label{intro}
\input{chapters/introduction}

\section{Related Work}\label{RW}
\input{chapters/related_work.tex}

\section{Data and Methodology}\label{method}
\input{chapters/data.tex}

\input{chapters/methodology.tex}

\begin{figure}[h!]
    \centering
    \begin{subfigure}[b]{0.38\linewidth}     
        \centering
        \includegraphics[width=\textwidth]{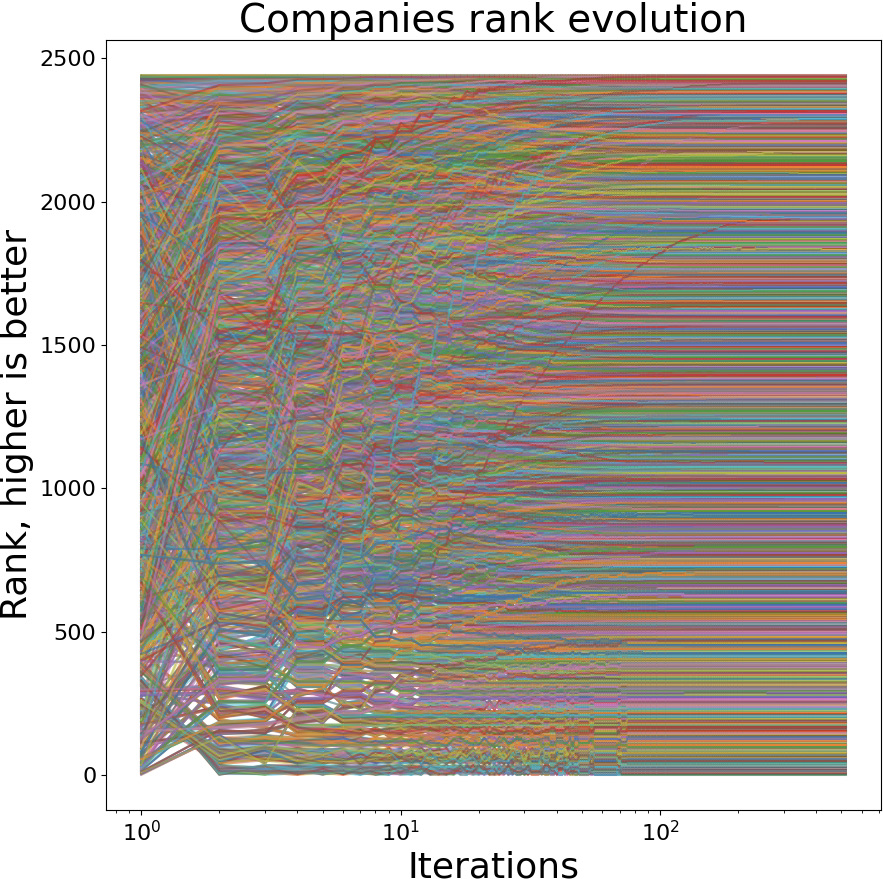}
    \end{subfigure}
    \hspace{1cm}
    \begin{subfigure}[b]{0.38\linewidth} 
        \centering
        \includegraphics[width=\textwidth]{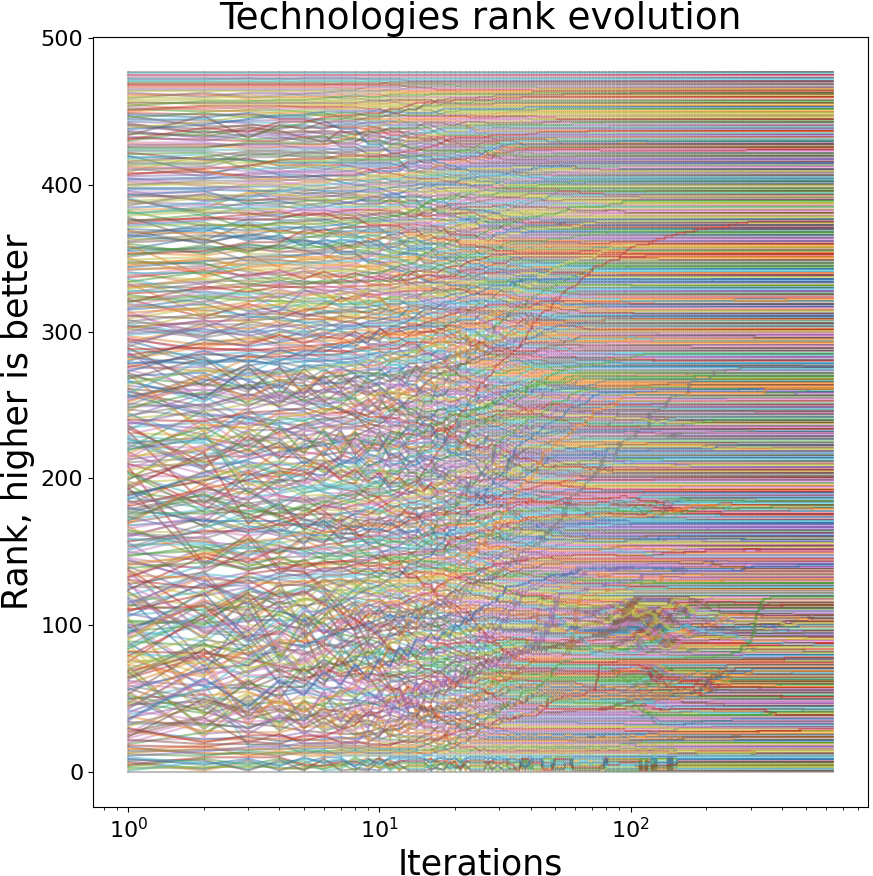}
    \end{subfigure}
    \hfill
    \caption{Convergence evolution of the rank algorithm.}
    \label{rank_evolution}
\end{figure}

\begingroup
\let\clearpage\relax
\endgroup
\section{Preliminary Results}\label{results}
\input{chapters/results}

\section{Further Steps}\label{further}

\input{chapters/further}

\section{Conclusion}\label{concl}
\input{chapters/conclusion}

%
%
%
\bibliographystyle{splncs04}
\bibliography{bib}

\end{document}

%% file: chapters/abstract.tex
The cybersecurity technological landscape is a complex ecosystem in which entities -- such as  companies and technologies -- influence each other in a non-trivial manner. Measuring the influence between entities is a tenet for informed technological investments in critical infrastructure. To study the mutual influence of companies and technologies from the cybersecurity field, we consider a bi-partite graph that links both sets of entities. Each node in this graph is weighted by applying a recursive algorithm based on the method of reflection. This endeavor helps to measure the impact of an entity on the cybersecurity market. Our results help researchers measure more precisely the magnitude of influence of each entity, and allows decision-makers to devise more informed investment strategies, according to their portfolio preferences. Finally, a research agenda is suggested, with the aim of allowing tailor-made investments by arbitrarily calibrating specific features of both types of entities.
 

%% file: chapters/introduction.tex
The operational continuity of critical infrastructure (CI) is central to a functioning modern society. However, CIs are managed and monitored by interdependent information systems, exposing CIs to cascading failures \cite{hines2007}. In this high-risk environment, the security of information systems is crucial \cite{dpd_PhD_2020}.\footnote{The hack of the Colonial Pipeline in May, 2021 is one of the most significant cyberattacks on a national CI in history. This case illustrates the high-risk environment CIs operate in and hence the importance of making informed technological investments that improve the (cyber) security of this type of infrastructure. Source: \url{https://www.bbc.com/news/technology-57063636}} However, developing and implementing an effective defence strategy to protect information-systems is a challenging task. The cybersecurity market is continuously exposed to complex and fast-paced technological developments. Research and development in the field is a costly and risky endeavour that leads to sunk costs and poor development prospects. Hence, uncertainties related to cybersecurity technologies -- and companies that develop them -- are ubiquitous \cite{anderson_why_2001}. Statistics point to a start-up failure rate of 90\%. In 42\% of cases, such failures are due to a misreading of market-demand, while in 29\% of cases, startups fail because they run out of funding \cite{fail_startups}. The decision-making process associated with cybersecurity investment and procurement has to work in this uncertain environment, which makes it hard to achieve high returns on investment. 
	
Our objective is to help CI decision-makers to make more informed investments when scouting and acquiring in the cybersecurity industry. For that purpose, we first model and map the ecosystem of entities (i.e., technologies, companies) from \textit{Crunchbase}. This mapping reflects the cybersecurity market through a bi-partite network. Then, we evaluate their relative influence in the whole ecosystem by adapting a recursive algorithm that returns a network-centrality measure. This should help decision-makers and investors quantitatively assess the influence of each entity in the cybersecurity ecosystem, thus reducing potential investment uncertainties and optimizing the procurement process.


The remainder of this paper will be structured as follows: Section \ref{RW} presents the related work; Section \ref{method} presents the data and methods; Section \ref{results} shows the preliminary results; Section \ref{further} sets the agenda for future work and discusses the limitations on current research; Finally, Section \ref{concl} concludes the whole.

%% file: chapters/related_work.tex
Network-centrality measures -- i.e., measures that assess the importance/influence of nodes in networks --  have been widely investigated (see \cite{battiston2020networks,landherr2010critical} for extensive literature reviews in the field of social networks and complex systems). Some research has focused on extensions and re-adaptations of the well-know Google \textit{PageRank} \cite{pagerank} algorithm, which also includes a bi-partite network to improve its indexing capability. For instance, Hidalgo et al. \cite{Hidalgo10570} developed the reflection method: an algorithm that characterizes the structure of bi-partite graphs by iteratively calculating the mean value of previous-step properties of a node's neighbors. Similarly, previous work by Klein at al. \cite{Maillart} assessed the relationship between editors and articles on Wikipedia by extending the \textit{PageRank} centrality measure. They developed a recursive algorithm to measure how editor expertise influences the quality of articles, and how contributions to articles influence editors' expertise. Their work shed light on cooperation and coordination dynamics, and the study of these systems has helped better solve common interaction problems that emerge within such social structures. 

To the best of our knowledge, despite a prolific literature related to economic valuation of cybersecurity investments (see \cite{schatz2017economic} for an extensive literature review), little work has tried to assess the importance/influence of entities in a network, especially in the field of cybersecurity and CIs. The idea of measuring the global rank of a node based on local information is therefore still an open research gap. This work tries to use an approach similar to Klein et al. to investigate the following research question: How does an entity's influence, predicated on its relations with other entities, allow for more informed investment strategies?

\if
The evaluation of startups and new technologies currently largely depends on investors' personal choice. However, Zhong et al. \cite{porfoliostrategy} highlight that the recent venture capital boom requires quantitative methodologies of screening and evaluation. They recognize that the birth of recent (tech) companies needs methodological decision making.
Previous work \cite{porfoliostrategy} does not leverage graph theory techniques, but proposed a two steps methodology that we also follow. First, the identification of the best entities to invest in and, then, the identification of the best investment strategy.
\fi

%% file: chapters/data.tex
We use a Crunchbase\footnote{\url{https://www.crunchbase.com/}; data downloaded on April 28th, 2021.} (CB) dataset to build our bi-partite network, which is composed of two node types : companies and technologies. 
CB encompasses information about company activity and financing by leveraging big data and open-source information in a semi-automated fashion. Data is sourced from investors and the community of contributors. 
As emphasized by Dalle et al. \cite{CBecomana}, CB is widely used by researchers because of the quality of its data and the usability of the platform. 
For the sake of brevity, we refrain from a detailed dataset description which can be found on the CB enterprise API. \footnote{\url{https://app.swaggerhub.com/apis-docs/Crunchbase/crunchbase-enterprise_api/}}

%% file: chapters/methodology.tex
\begin{figure}[b!]
    \centering
    \includegraphics[scale=0.25]{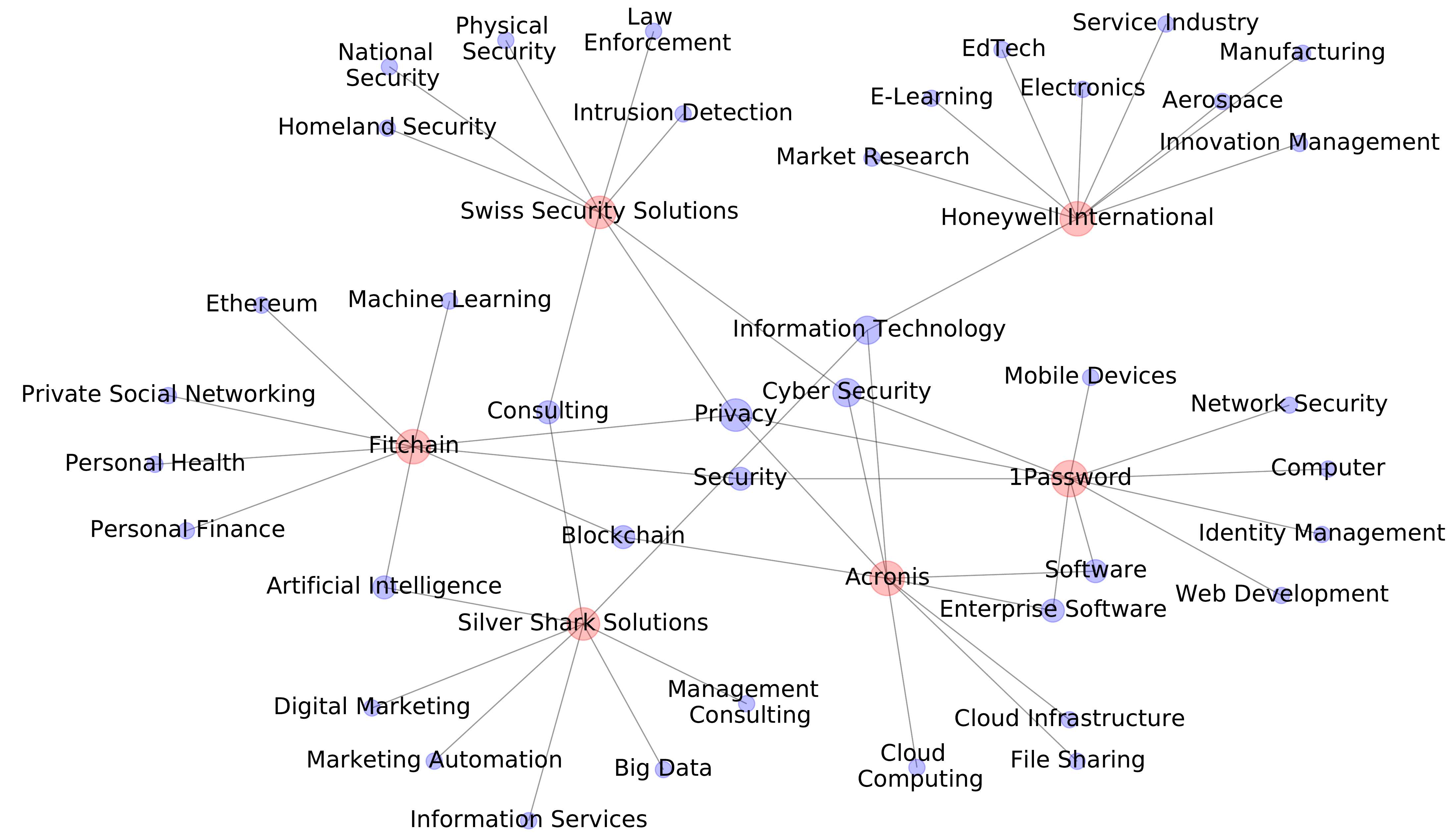}
    \caption{Bi-partite network for selected cybersecurity companies (red nodes) and technologies (blue nodes) they are working on. The nodes' size represents the number of neighbors.}
    \label{net}
\end{figure}

Figure \ref{net} gives an idea of the structure of the bi-partite network that describes which technologies each company is working on.

We adapt the recursive algorithm developed by Klein at al. \cite{Maillart}, based on the method proposed by Hidalgo et al. \cite{Hidalgo10570}. This new method models the complex structure of cooperation and competition occuring in the cybersecurity space. The algorithm produces a ranking of technologies and companies from most to least influential. This ranking thus condenses the positive influence of experienced companies on technologies as well as the positive impact of newborn companies on novel fields. In the same way, important technologies will positively influence companies that are linked to them, and this process iteratively increases the rank of both the companies and the technologies. Moreover, in their paper on Wikipedia editorial work, Klein et al. \cite{Maillart} claim that too many editors working on an article can sink the value of the entities. Our network analysis also investigates this phenomenon of negative influence in the context of cybersecurity technologies. This would mean for instance that if too many companies work on the same technology, the business gap narrows, and companies lose market share. This indisputably reduces a company's value.  

Starting from Klein at al. \cite{Maillart}, we build  an adjacency matrix $M_{c,t}\in \mathbb{R}^{N_c, N_t}$ that takes value 1 if a company $c$ works on a technology $t$ and 0 otherwise. $N_c$ represents the total number of cybersecurity companies we are considering and $N_t$ the total number of technologies. The aim of the algorithm is to assign a weight to every node, which sums up its relevance within the graph: a contribution value for each company and a quality value for each technology. The starting point consists in measuring the expertise of a company ($w_c^{0}$) by summing up the number of technologies it works on: $w_c^{0} = \sum_{t=1}^{N_t} M_{c,t} = k_c$. The same holds for technologies: $w_t^{0} = \sum_{c=1}^{N_c} M_{c,t} = k_t$.

First, we recall that this algorithm is a Markov process: the step $w^{n}$ depends only on information available at the previous step $w^{n-1}$. At each step, the method incorporates information about the expertise of companies and the relevance of technologies, leveraging the bi-partite network structure. The whole process can be seen as a random walker that jumps with a transition probability that is zero in case $M_{c,t}=0$. We need to define two matrices that explain how we move from one step to another one: they represent the probability of jumping from technology $t$ to company $c$ and depends on the initial conditions.
\begin{equation}\label{step_0}
    \begin{cases}
    G_{c,t}(\beta) = \frac{M_{c,t}k_c^{-\beta}}{\sum_{c'=1}^{N_c} M_{c',t}k_c'^{-\beta}}  \\
    G_{c,t}(\alpha) = \frac{M_{c,t}k_t^{-\alpha}}{\sum_{t'=1}^{N_t} M_{c,t'}k_t'^{-\alpha}}
    \end{cases}
\end{equation}
Klein at al. \cite{Maillart} also introduce two parameters, $\alpha$ and $\beta$, that measure how coordination generates value. Thanks to $G_{e,a}$, we get the recursive step:
\begin{equation}
    \begin{cases}
    w_c^{n+1} = \sum_{t=1}^{N_t} G_{c,t}(\beta) w_c^{n}  \\
    w_t^{n+1} = \sum_{c=1}^{N_c} G_{c,t}(\alpha) w_t^{n}
    \end{cases}
\end{equation}
Similarly to \textit{PageRank}, the recursion ends when the rank stabilizes. 

The CB platform assigns a rank to the top companies -- according to their algorithm -- in each industry. The CB rank takes into account the entity's strength of relationships, funding events, news articles, acquisitions, etc.\footnote{\url{https://about.crunchbase.com/blog/influential-companies/}}. We compare our results in cybersecurity with this rank. We investigate the strength and direction of the association between the two scores using Spearman's rank correlation coefficients.

%% file: chapters/results.tex
Our first results are based on the selection of all companies whose description contains at least two words related to the field of cybersecurity.  We get a total of 2,443 companies and 478 technologies. Figure \ref{rank_evolution} shows that the recursive algorithm introduced in \autoref{method} converges for both companies and technologies after a sufficient number of iterations (421 and 538 respectively). 

As mentioned in \autoref{method}, we compare our ranking with the CB rank, hereafter designated as the baseline. Thus, to make the ranks comparable, we convert our algorithm's output into a ranking. The resulting Spearman's correlation (0.014) shows that the two ranks are not correlated: even if the goals of the \textit{TechRank} and the CB rank are similar, this outcome reflects their substantial differences. First, we do not know the exact mechanism by which CB (which is not open source) ranks the entities, which offers little information to investors trying to make investments according to technology or market preferences. Moreover, the CB score focuses more on the level of activity of the company, rather than its influence on the market. Among all the factors that influence the CB ranking, nothing is said about the influence that technologies have on the company's value and market importance. Moreover, CB simply ranks companies, while our algorithm assigns a weight, which gives not just an ordering of the companies, but also gives a quantitative measure of the distance between one entity and the next one. These discrepancies lead to different results and we believe that our less-is-more approach, which directly depends on the capabilities of the companies, is a good way for investors to analyze an entity's development prospects. This, together with the next steps explained in \autoref{further}, could lead to a personalisable and transparent approach to company rankings.  

Another benefit of our methodology is the technology ranking, which allows investors to make portfolio decisions not just based on company insights, but also on technology insights. The \textit{TechRank} methodology thus enables them to create the portfolio that better reflects their preferences. Finally, a last finding of the research points to the fact that new fields have a good impact on companies working on them and that too much competition on a technology has a negative impact on its neighbors, thus confirming our secondary hypothesis. 

%% file: chapters/further.tex
Our research agenda will focus on two steps: first, including the influence of exogenous factors -- such as impact of incubators and social aspects-- and second, creating the optimal portfolio strategy. The first step incorporates the impact of previous investments, captured by another bi-partite structure composed of investors and companies: each company is linked to its investors and edges are weighted by the investment amount. Tracking the investments is a relevant factor for a \say{follow the money} strategy and describes the investor trust in the company. Once we have defined the entities to invest in, for the last step, we use modern portfolio theory to maximise returns while minimizing variance. The final outcome will allow investors to personalize the algorithm according to their preferences thanks to a transparent a customisable platform.

%% file: chapters/conclusion.tex

The aforementioned methodology constitutes the first step towards a new data-driven investment strategy, which gives investors a succint and easily understable ranking of companies and technologies based on their influence in the market. 

Thanks to the interdisciplinary core of this solution, investors can undertake a transparent decision making process when dealing with highly complex scenarios, such as in the cybersecurity market. In finance, the efficacy of both “classical” technical and fundamental analysis is disputed by the efficient-market hypothesis. Our \textit{TechRank} is the kickoff for a complementary (or even alternative) modern technological portfolio analysis for CI operators. We believe also that the algorithm is extendable to every large organization dealing with a high level of uncertainty. Therefore, we expect that the \textit{TechRank} algorithm will be applied and further developed in other fields.